**Strong influence of boron precursor powder on the critical current density of MgB$_2$**

S. K. Chen, K. A. Yates, M. G. Blamire and J. L. MacManus-Driscoll

Department of Materials Science and Metallurgy, University of Cambridge,

Pembroke Street, Cambridge, CB2 3QZ, UK.

**Abstract**

The influence of the nature of the boron precursor on the superconducting properties of polycrystalline MgB$_2$ was studied. Critical current densities ($J_c$'s) for the MgB$_2$ made from high purity amorphous boron are at least a factor of three higher than typical values measured for standard MgB$_2$ samples made from amorphous precursors. Two possible mechanisms are proposed to account for this difference.

Samples made from crystalline boron powders have around an order of magnitude lower $J_c$'s compared to those made from amorphous precursors. X-ray, $T_c$ and resistivity studies indicate that this is as a result of reduced current cross section due to the formation of (Mg)B-O phases. The samples made from amorphous B contain far fewer Mg(B)-O phases than crystalline B despite the fact that the amorphous B contains more B$_2$O$_3$. The different reactivity rates of the precursor powders accounts for this anomaly.

## 1. Introduction

Recently, there has been a considerable amount of research undertaken in order to understand the influence of dopants and nanoparticle additions to enhancing the $J_c$ of the $MgB_2$ [1-5]. However, before understanding how these additions influence $J_c$, it is important to understand how intrinsic impurities in $MgB_2$ (in particular oxides of Mg and B constituents, and Mg stoichiometry) enhance or degrade $J_c$. There has been much discussion of the difference between 'clean' and 'dirty' 'undoped' samples, and about difficulties in relating sample resistivity to $T_c$ and $J_c$. Clearly, there are many factors involved including reduction of current carrying cross section as a result of intergrain second phases, formation of grain boundary Josephson junctions by grain boundary phases, intragrain scattering as a result of poorly controlled stoichiometry (e.g. Mg vacancies, oxygen substitution, or possibly even B vacancies), or scattering from intragrain nano-scale precipitates [6, 7]. Control of these scattering mechanisms is at an early stage and is key for understanding the extent to which doping of intragranular regions (e.g. with C) may also unintentionally produce deleterious intergrain second phases.

The aim of this study is to understand how the $J_c$ versus field behaviour for $MgB_2$ is influenced by the form and purity of the boron precursor. In addition to the scientific interest of the work, the study has considerable merit from the application point of view.

## 2. Experimental details

MgB$_2$ samples were prepared by reacting Mg and B using conventional solid state reaction. Crystalline magnesium powder (Alfa Aesar, 99.8 %, 325 mesh) and four different types of boron powders (2 different purities of both crystalline B and amorphous B) as shown in table 1. Mg and B were mixed in the stoichiometric ratio of 1:2, followed by grinding for 1 h. The resultant mixture was then pressed into pellets 5 mm in diameter and about 2 mm thick using a hydraulic press with an applied load of 2 tonnes. The pellets and approximately the same weight fraction of Mg curls (~0.003g per 0.030g 'Mg+2B') were wrapped in Ta foil, placed inside a tubular furnace and annealed at 900 ºC for 15 min in a reducing atmosphere of 2% H$_2$-Ar. The heating and cooling rates used were 15 ºC/min. Several different samples were made in separate reaction runs to check for reproducibility.

X-ray diffraction in the $\theta$ - $2\theta$ step-scanning mode with 0.05º increments were recorded in a Philips PW1050 diffractometer with a Cu-K$_\alpha$ radiation source. Room temperature resistivity was measured using a standard four probe technique. In order to eliminate the effect of Mg which had condensed on the surface of some of the pellets during annealing, each sample was polished before the measurement. The superconducting transition temperature, $T_c$, was obtained using a commercial Quantum Design DC Magnetic Properties Measurement System (MPMS) by first cooling the sample in zero field and then measuring the magnetic moment as the sample was warmed in field. Magnetisation hysteresis loops were performed on bar shaped samples of ~2 mm$^3$ volume with the magnetic field applied parallel to the longest dimension of the sample. Magnetic critical current density was estimated based on the critical state model [8].

Particle size distribution was determined using a Mastersizer E Particle Size Analyser from Malvern Instruments. To ensure reliable data acquisition, 0.05g of boron powder was ultrasonically dispersed in acetone to avoid agglomeration of powder particles. The equipment was calibrated each time before the subsequent measurement to reduce the background spectrum.

## 3. Results and discussion

Figure 1 shows the x-ray diffractogram for the crystalline boron (figure 1(a)) and amorphous boron powders (figure 1(b)). The patterns can be indexed according to single phase boron, except for the $B_2O_3$ peaks indicated with dashed lines. There is also an unidentified peak at around 18.4°. It is difficult to quantify the differences in the amount of $B_2O_3$ between the different B's, e.g. the $B_2O_3$ peaks appear high in the amorphous boron powder (B-A9999) but this is set against the low intensity amorphous B features. Nevertheless, it is expected that the amorphous powders contain more $B_2O_3$ since they are fabricated by high temperature Mg reduction of $B_2O_3$. It is interesting that the B-A9597 powder shows some crystalline peaks from B, despite being labelled as amorphous.

The crystalline B is expected to contain less oxygen impurities since it is formed by high temperature reduction of halides of B with $H_2$ gas. For the amorphous boron, the difference between different purity powders relates to the amount of excess Mg present as a result of the reduction.

X-ray diffractograms of the $MgB_2$ pellets made from the corresponding B precursor powders are shown in figure 2. MgO was present as a second phase in measurable quantities in *all* the $MgB_2$ samples, although it has the smallest fraction in the purest amorphous B precursor sample, A9999. The samples formed from the amorphous powders only showed the presence of MgO as a second phase.

The impurities in the crystalline (C) precursor samples consisted of boron oxides and magnesium boron oxides. In addition, several peaks due to unidentified phases were clearly observed, and are marked with circles. Some un-reacted Mg was also found in sample C98 (excess Mg was always found in samples made from the B-C98 batch).

Summarising the x-ray data, more $(Mg)B_xO_y$ impurity phases were formed in samples prepared from the crystalline rather than amorphous boron. At first, this finding seems surprising considering that the crystalline B precursor powders are expected to contain less oxygen, and that they contained less $B_2O_3$ than the amorphous powders (figure 1). The results can be explained based on the reactivity of the powders. Table 1 shows that the crystalline boron contains large particles of 10's of micron in size, whereas the amorphous powders have particle sizes of ~ 0.5 µm. The reactivity of the amorphous powders is much greater than the crystalline powders, and the reduced particle size further enhances the reaction rate [9]. Hence, when the crystalline B powders come into contact with surface oxidised Mg powder, the reaction to form $MgB_2$ is relatively slow and there is sufficient time for $(Mg)B_xO_y$ phases to form. On the other hand, for the amorphous powder the Mg reacts rapidly with B to form $MgB_2$ and with $B_2O_3$ to form MgO+B.

Table 2 lists the lattice strain estimated from the Williamson-Hall plot [10]. The peak breadth and $2\theta$ values were refined and carefully checked to avoid overlapping from impurity phases. Comparison of the experimental data with a Si standard also showed that size broadening effects were small. Within the error of calculation, there was no significant difference in relative strain amongst samples and there was no evidence of anisotropic strain. The cell parameters were calculated using the Rietveld method in the approximation of a pseudo-voigt function by taking into account the contribution from thermal displacement of atoms. The hexagonal crystal structure of $MgB_2$ [11] was used as the reference data and the MgO phase was included in the refinement. The *a*-axis for A9597 was the smallest amongst the samples while the *c*-axis did not change to within error. Small '*a*' values have previously been related to intergranular strain effects [12] and this possibility is discussed for sample A9597 in relation to the observed microstructure. Nevertheless, the very small differences in '*a*' between samples indicates that there are no significant structural perturbations within the crystallites.

Typical SEM images of the four different samples are shown in figure 3. The larger grains of the crystalline samples (few hundred nm) compared to the amorphous samples (~ 100 nm and less) is consistent with the measured sizes of the precursor boron powders of table 1. In addition, the bimodal particle sizes observed in the B-C99 and B-A9597 boron precursor powders (table 1) are also reflected in bimodal grain sizes in the resulting $MgB_2$ samples, particularly for the A9597 sample. This strong bimodality in A9597 likely leads to greater intergranular strain, and thus accounts for smaller '*a*'. The A9999 sample has the finest and most uniform grain size.

Figure 4 shows the superconducting transition temperature measured at 20 Oe. $T_c$ was determined by taking the first deviation point from linearity that signifies the transition from the normal to superconducting state. The $T_c$'s were in the range of 37.9K (C99) to 38.8K (C98). Both A9597 and A9999 showed the same $T_c$ of 38.2K. Our results are in contrast to Ribeiro *et al* who found that $T_c$ is improved with boron purity [13]. The breadths of the $T_c$ transitions are larger for the crystalline precursor samples than for the amorphous precursor samples (i.e. 2 - 3 K compared to 0.75 – 1 K, respectively, using a 10 - 90% criterion) suggestive of greater sample inhomogeneity and/or poorer intergrain connectivity for the crystalline precursor samples.

Figure 5 compares the field dependence of $J_c$ up to 5T. The form of $J_c$ (*H*) was identical and $J_c$ values were reproducible to within 25% for the same precursors from one sample batch to the next. Higher $J_c$'s are achieved for the amorphous precursor samples, despite the lower densities (table 2) than for the crystalline precursor samples.

Both at 6K and 20K, $J_c$ ($H$) of A9999 is an order of magnitude larger than C99. C98 has the lowest $J_c$ of all the samples and $J_c$ decreased rapidly with $H$, again suggestive of poor intergrain connectivity for the crystalline samples. The $J_c$ ($H$) data is consistent with the x-ray data (figure 2), namely, that the crystalline precursor samples contained a significant amount of oxide impurities ($B_xO_y$, Mg-B-O, and MgO) whereas the amorphous precursor samples contained only MgO.

The high resistivity values for the crystalline precursor samples (~230 - 340 $\mu\Omega$.cm) suggests that the oxide phases decrease the current carrying cross section by obstructing some of the intragrain regions. From XRD, the main difference between C98 and C99 is presence of Mg in C98, as well as a larger amount of MgO. The fact that the resistivities for C98 are only marginally higher than for C99 (table 2) suggests a combined effect of a decrease in resistivity due to the presence of extra Mg, but also an increase due to extra MgO, some of which may come from a surface oxide layer on the Mg which is small in volume fraction (hence it does now show up strongly in x-ray) but large in surface area (hence leading to an increase in the resistivity).

The highest $J_c$ of all the samples was achieved in sample A9999 prepared using the highest purity boron. Zhou *et al* previously showed that samples made from different purity amorphous borons exhibited different $J_c$'s, and also that the highest purity boron (99.99%) yielded the highest $J_c$'s [14]. However, the values reported were lower than in this study e.g. by a factor of ~ 4-5 at 20K, 3T. As far as known, the only other report of similar high $J_c$ ($H$) values as reported here was again in $MgB_2$ [7] made using similar high purity amorphous boron. Despite the high purities of the precursors, the samples were believed to be rather

'dirty' because room temperature resistivity values were ~ 180 $\mu\Omega$.cm. Similar high resistivities were also measured in our samples, as discussed below.

Considering that the A9999 samples showed similar x-ray diffraction patterns to the A9597 samples, except for a slightly larger fraction of MgO in A9597, it might be expected that the $J_c$'s would be similar, and not a factor of 3 or more lower. The room temperature resistivity (~ 50 $\mu\Omega$.cm) of A9597 is typical for a clean $MgB_2$ sample and is around a factor of 5 lower than the crystalline precursor samples which contained several impurity phases. Surprisingly, the 'clean' A9999 samples which were also expected to have resistivities of ~50 $\mu\Omega$.cm or lower, similar to the A9597 samples, had high resistivities of ~ 200 $\mu\Omega$.cm, approaching the crystalline precursor samples. It is possible that these resistivity values were artificially high because the samples were brittle and sometimes cracked when contacts were applied to them – the reason for the brittleness can be attributed to the uniform, fine grain structure as shown in figure 3(d) but this would *not* explain the higher $J_c$ values of the A9999 samples. Two other possible reasons for the higher resistivities are: a) larger grain boundary contribution to resistivity because of the very fine grain size, or b) possible presence of Mg non-stoichiometry in the $MgB_2$ grains at a level insufficient to cause a degradation of $T_c$. In the former scenario, the $J_c$ enhancement in the A9999 samples would be by additional grain boundary pinning, and in the latter scenario $H_{c2}$ would be increased by enhanced intragrain scattering. Since $J_c$ (*H*) behaviour for the A9999 samples is, in fact, better than some previously reported doped samples [15], whatever is causing the significant enhancement in $J_c$ over normal amorphous precursor $MgB_2$ samples needs to be elucidated before a clear understanding is formulated about the influence of doping on $J_c$.

## 4. Conclusions

In conclusion, we have shown that the form and purity of boron precursors are of critical importance to the $J_c$ ($H$) behaviour of $MgB_2$ samples. The $J_c$ ($H$) behaviour of the samples made from the crystalline boron precursors can be explained in terms of a reduced effective cross sectional area as a result of the presence of several oxide impurities. The differences between amorphous boron powders are less clear, with very large differences in $J_c$ ($H$) obtained from samples showing very similar phase purities. Nevertheless, it is clear that to obtain reproducible $J_c$'s, very careful control of amorphous B starting powders and reaction conditions are required.

## Acknowledgements

We would like to thank Mary Vickers for her useful discussion. We acknowledge EPSRC for funding the research. S. K. Chen also acknowledges the financial support from UPM.

Table 1. Source, form and purity of the different boron powders with their particle size distribution.

| Boron powder | Source | Form | Purity | Peak value(s) of particle size distribution (μm) |
|---|---|---|---|---|
| B-C98 | Alfa Aesar | Crystalline | 98 % | 21.10 [325 mesh] |
| *B-C99 | FluoroChem | Crystalline | 99 % | 11.42, 0.56 [< 40 μm] |
| *B-A9597 | Fluka | Amorphous | 95 – 97 % | 0.56, 2.42 [ - ] |
| B-A9999 | Alfa Aesar | Amorphous | 99.99% | 0.54 [325 mesh] |

[ ]: Particle size as indicated in the chemical label.

*: Powders with dual particle size distribution.

Table 2. Room temperature resistivity, density, relative strain and cell parameters of samples prepared from different boron powders.

| Sample | Resistivity (μΩ.cm), $\rho$ [290 K] | | Density (g.cm$^{-3}$) | Strain (%) | Cell parameters (Å) | |
|---|---|---|---|---|---|---|
| | Pellet 1 | Pellet 2 | | | $a$ | $c$ |
| C98 | 274.8 | 344.1 | 1.48 | 0.171 ± 0.043 | 3.0826 ± 0.0007 | 3.520 ± 0.001 |
| C99 | 242.6 | 230.3 | 1.45 | 0.188 ± 0.046 | 3.0820 ± 0.0010 | 3.524 ± 0.002 |
| A9597 | 44.6 | 60.7 | 1.16 | 0.211 ± 0.031 | 3.0789 ± 0.0006 | 3.521 ± 0.001 |
| A9999 | 178.4 | 225.3 | 1.23 | 0.198 ± 0.019 | 3.0822 ± 0.0005 | 3.525 ± 0.001 |

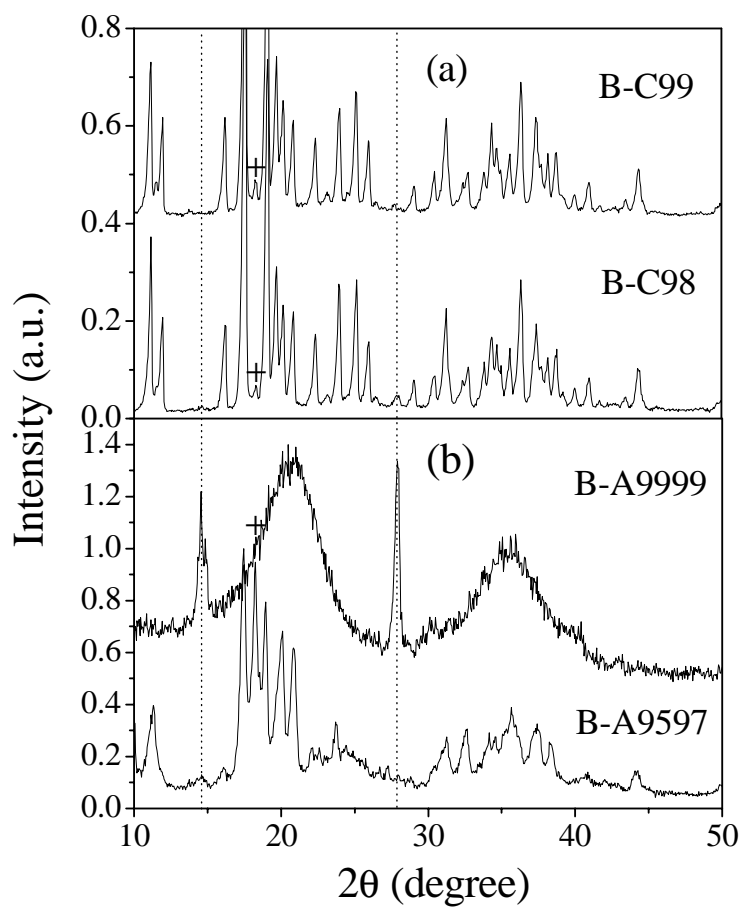

Figure 1. X-ray patterns of (a) crystalline boron powder and (b) amorphous boron powders. The dashed lines indicate the positions of two of the $B_2O_3$ peaks. + denotes unidentified impurity.

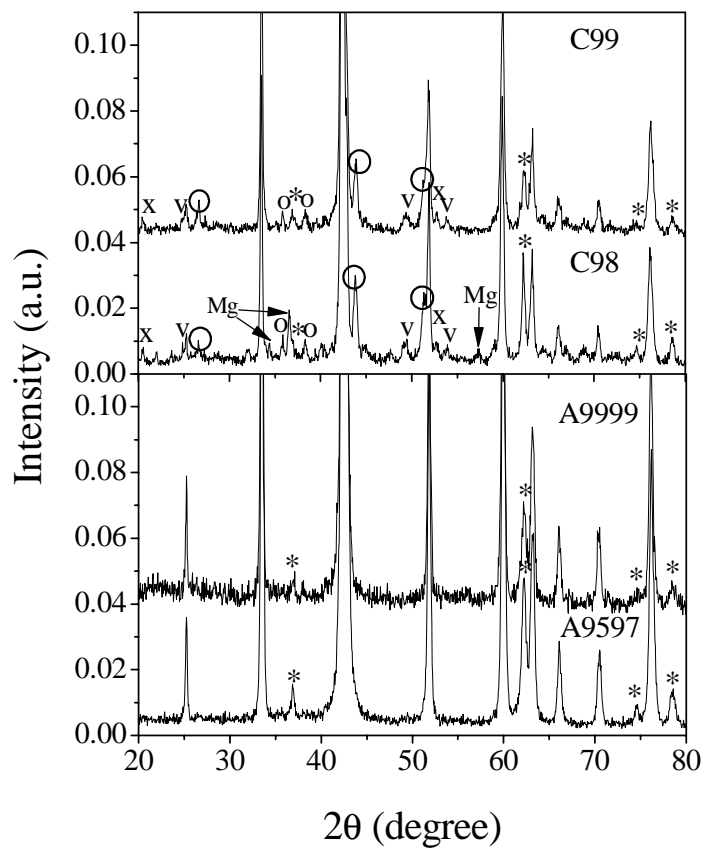

Figure 2. X-ray powder diffraction patterns of MgB$_2$ samples made from different B powders. The unlabelled peaks are MgB$_2$. X: B$_{13}$O$_2$; V: Mg$_3$(BO$_3$)$_2$; o: B$_2$O; *: MgO.

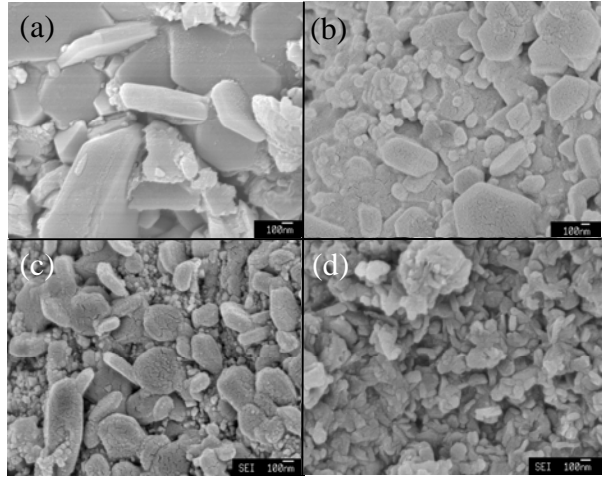

Figure 3. FEG-SEM images of samples (a) C98 (b) C99 (c) A9597 and (d) A9999.

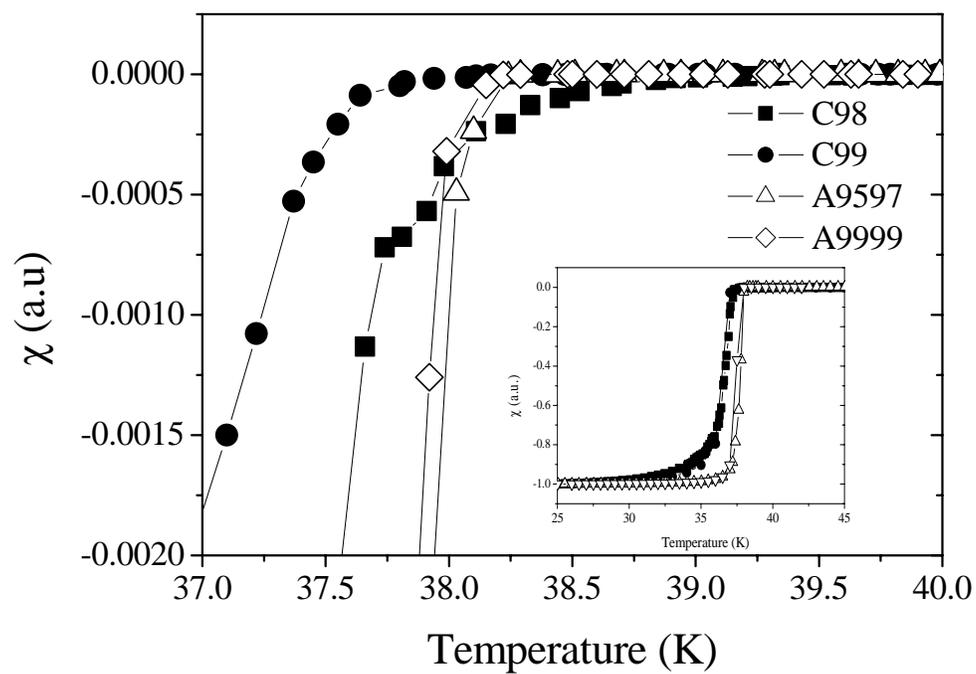

Figure 4. Temperature dependence of normalised susceptibility. Inset: transition temperature in a wider temperature range.

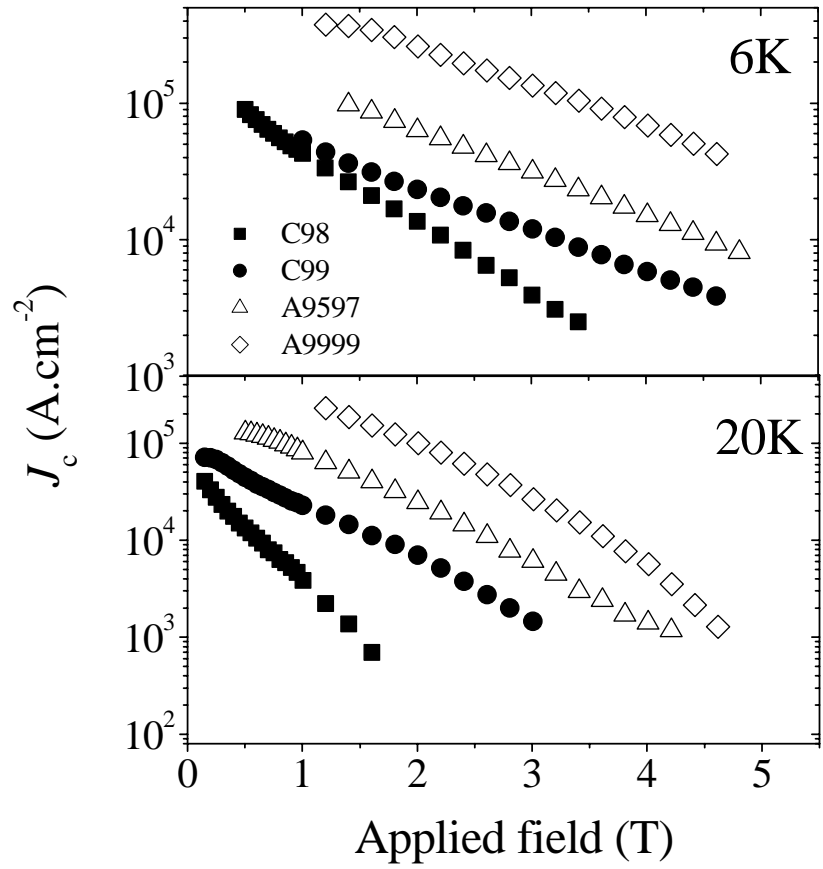

Figure 5. Magnetic critical current densities versus applied magnetic field at 6K and 20K.